\documentclass[aps,twocolumn,preprintnumbers,superscriptaddress,prb]{revtex4-2}
\usepackage{graphicx}
\usepackage{amsmath}
\usepackage{amssymb}
\usepackage{color}
\usepackage{float}
    
\begin{document}

\title{Topological Transitions with an Imaginary Aubry-Andr\'e-Harper Potential}

\author{Bofeng Zhu}

\affiliation{Division of Physics and Applied Physics, School of Physical and Mathematical Sciences,\\
Nanyang Technological University, Singapore 637371, Singapore}

\affiliation{School of Electrical and Electronic Engineering, \\
Nanyang Technological University, Singapore 637371, Singapore}
  
\author{Li-Jun Lang}

\affiliation{Guangdong Provincial Key Laboratory of Quantum Engineering and Quantum Materials,\\
School of Physics and Telecommunication Engineering, South China Normal University, Guangzhou 510006, China}

\author{Qiang Wang}

\affiliation{Division of Physics and Applied Physics, School of Physical and Mathematical Sciences,\\
Nanyang Technological University, Singapore 637371, Singapore}

\author{Qi Jie Wang}

\email{qjwang@ntu.edu.sg}

\affiliation{Division of Physics and Applied Physics, School of Physical and Mathematical Sciences,\\
Nanyang Technological University, Singapore 637371, Singapore}

\affiliation{School of Electrical and Electronic Engineering, \\
Nanyang Technological University, Singapore 637371, Singapore}
  
\affiliation{Centre for Disruptive Photonic Technologies, Nanyang Technological University, Singapore 637371, Singapore}

\author{Y.~D.~Chong}

\email{yidong@ntu.edu.sg}

\affiliation{Division of Physics and Applied Physics, School of Physical and Mathematical Sciences,\\
Nanyang Technological University, Singapore 637371, Singapore}

\affiliation{Centre for Disruptive Photonic Technologies, Nanyang Technological University, Singapore 637371, Singapore}

\begin{abstract}
  We study one-dimensional lattices with imaginary-valued Aubry-Andr\'e-Harper (AAH) potentials.  Such lattices can host edge states with purely imaginary eigenenergies, which differ from the edge states of the Hermitian AAH model and are stabilized by a non-Hermitian particle-hole symmetry.  The edge states arise when the period of the imaginary potential is a multiple of four lattice constants.  They are topological in origin, and can manifest on domain walls between lattices with different modulation periods and phases, as predicted by a bulk polarization invariant.  Interestingly, the edge states persist and remain localized even if the gap in the real spectrum closes.  These features can be used in laser arrays to select topological lasing modes under spatially extended pumping.
\end{abstract}

\maketitle

\section{Introduction}

The Aubry-Andr\'e-Harper (AAH) model is a foundational theoretical model that illustrates the deep connections between quasicrystals, localization, and band topology \cite{Harper1955, Aubry1980, Thouless1982}.  It consists of a one dimensional (1D) periodic discrete lattice, on which is applied a sinusoidal potential with a mismatched period.  Varying the potential's period and phase produces an assortment of spectral gaps, which map to the band gaps of a two dimensional (2D) quantum Hall lattice \cite{tknn, Janot1994, Kraus2016, Zilberberg2021}.  The boundary states in certain gaps of the 1D AAH model likewise map to topological edge states of the 2D lattice, which are linked to bulk topological invariants (Chern numbers).  These interesting features have inspired numerous investigations into variants of the AAH model.  For example, an AAH-type model with commensurate hopping modulations was found to have a separate class of topological boundary states \cite{Ganeshan2013}: zero modes whose energies are pinned to zero by particle-hole symmetry \cite{Ryu2002} and are linked to the topological properties of the Majorana chain \cite{Kitaev2001}.

Over the past decade, there has been increasing interest in \textit{non-Hermitian} extensions of the AAH model \cite{Longhi2014, Yuce2014, Harter2016, Longhi2019, Jiang2019, Zeng2020, TLiu2020, YLiu2020, Longhi2021, Weidemann2022}, as part of a broader program to explore the properties and uses of non-Hermitian systems \cite{Ozdemir2019, Kawabata2019, Ashida2020}.  These models have included AAH-type lattices with parity/time-reversal (PT) symmetric gain/loss \cite{Longhi2014, Yuce2014, Harter2016, Longhi2019, YLiu2020}, and lattices with asymmetric hoppings violating both Hermiticity and reciprocity \cite{Jiang2019, Longhi2021, Weidemann2022}.  For example, PT symmetric AAH models have been found to exhibit fractal spectra, similar to the Hermitian AAH model, in the real part of their eigenenergies \cite{Yuce2014}.  Their PT symmetry breaking transition points also have interesting properties \cite{Harter2016, Longhi2019, YLiu2020}, such as governing the formation of boundary states \cite{Harter2016} and mobility edges \cite{YLiu2020}.

The boundary states in these non-Hermitian AAH models are directly related to the boundary states of the original AAH model.  Similar persistence of topological boundary states into the non-Hermitian regime has been observed in other models; for example, in PT symmetric Su-Schrieffer-Heeger (SSH) models \cite{Schomerus2013, Lang2018, Pan2018, Parto2018, Zhao2018}, topological zero modes can be stabilized by particle-hole symmetry (as in the original Hermitian SSH model) or a non-Hermitian particle-hole symmetry \cite{Ge2017,Qi2018}.  Very recently, researchers have also found lattice models that host intrinsically non-Hermitian boundary states with no direct link to the Hermitian case \cite{Lee2016, Leykam2017, Takata2018}.  For instance, Takata and Notomi discovered a periodic 1D lattice, with four atoms per unit cell, that hosts zero modes induced purely by gain and loss \cite{Takata2018}.  In view of these advances, it is worthwhile to examine zero modes in non-Hermitian AAH models.  Can such modes be induced by gain/loss?  What topological properties govern them, and how are they influenced by the AAH-style potential?

\begin{figure*}
  \centering
  \includegraphics[width=0.98\textwidth]{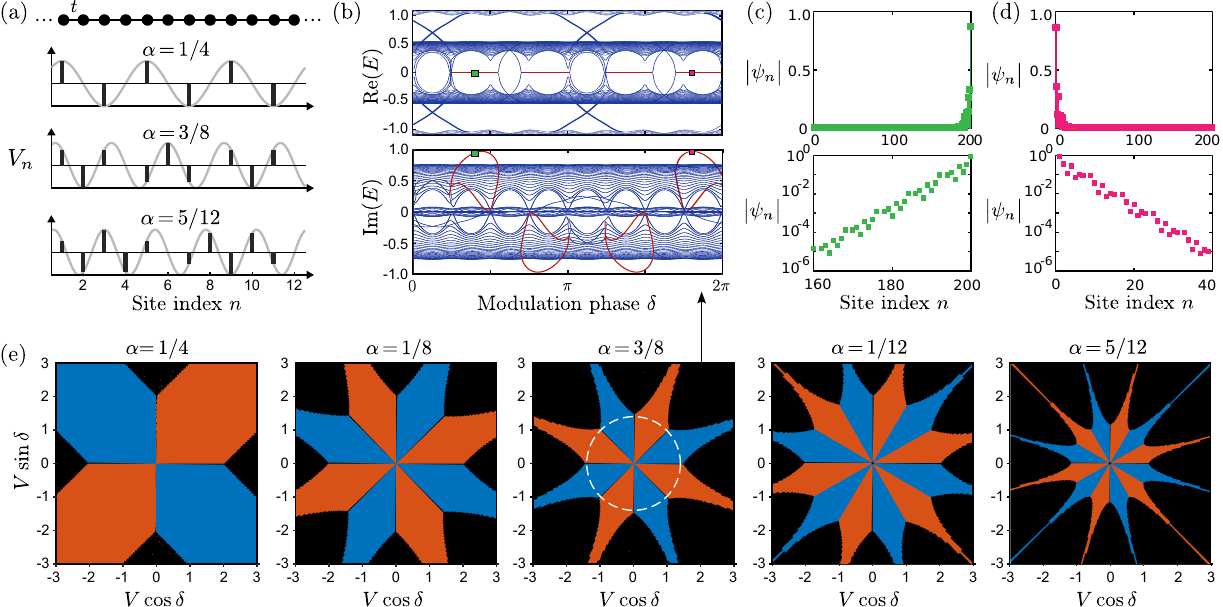}
  \caption{(a) Schematic of a one-dimensional lattice with imaginary modulation.  The model consists of a discrete chain with uniform nearest neighbor hopping $t$ (top panel), with imaginary on-site mass $iV_n$ varying sinusoidally in space, with wavenumber $2\pi \alpha$.  The bottom panels plot $V_n$ versus site index $n$ for $\alpha = 1/4$, $3/8$, and $5/12$ (black bars), along with the modulation profile (light gray curves).  (b) Plot of the complex eigenenergies $E$ versus modulation phase $\delta$, for a finite lattice of size $N = 200$ with open boundary conditions.  The model parameters are $V=1.4$ and $\alpha=3/8$, corresponding to the white dashes in (e).  The zero modes are plotted in red.  (c)--(d) Wavefunction magnitude $|\psi_n|$ versus site index $n$ for the zero modes at (c) $\delta = 0.4\pi$ and (d) $\delta = 1.8\pi$, respectively indicated by the green and pink dots in (b).  Lower panels are semi-logarithmic plots for the sites nearest to one lattice boundary, showing that the zero modes are exponentially localized at the boundary.  (e) Phase diagrams for different $\alpha$, with orange (blue) regions indicating gaps at $\mathrm{Re}(E)=0$ that are topologically nontrivial (trivial) according to the polarization and global Berry phase invariants calculated from the bulk band structure \cite{SM}.  In the black regions, the real part of the bulk spectrum is gapless at $\mathrm{Re}(E)=0$.}
  \label{fig:schematic_phase}
\end{figure*}

Here, we investigate a non-Hermitian AAH model with imaginary commensurate potentials.  We find that when the modulation has period $\lambda = p/q$, where $p$ and $q$ are coprime integers and $p$ is a multiple of 4, there arise topological boundary states whose energies have zero real part, which we refer to as ``zero modes''.  The case of $\lambda = 4$ corresponds to the Takata-Notomi lattice \cite{Takata2018}.  The zero modes are stabilized by a non-Hermitian particle-hole symmetry \cite{Ge2017,Qi2018}, and are linked to a non-Hermitian topological invariant based on the electric polarization \cite{Luo2019, LeePRB2020, Taberner2022, Hu2022}, which depends on the modulation parameters.  We derive the topological phase diagrams, and show that they predict the existence of zero modes at domain walls between different modulation functions (including those with different periods).  Interestingly, the zero modes can survive and retain their localized character even if the gap in the real spectrum closes.

As the zero modes are governed by an imaginary sinusoidal potential, it may be possible to use them for mode selection in laser arrays.  In existing implementations and proposals for topological lasers \cite{StJean2017, Ota2018, Parto2018, Zhao2018, Harari2018, Bandres2018, ZengYQ2020, ZYang2020, Harder2021, Dikopoltsev2021, Choi2021, Qi2019}, including those based on the 1D SSH lattice \cite{StJean2017, Ota2018, Harder2021} or its PT symmetric variant \cite{Parto2018, Zhao2018}, it is typically necessary to selectively pump the spatial regions where the desired topological modes are localized \cite{StJean2017, Parto2018, Zhao2018, Harari2018, Bandres2018, Ota2018, ZengYQ2020, ZYang2020, Harder2021, Dikopoltsev2021, Choi2021}. This induce the topological modes, rather than the numerous other non-topological modes, to lase.  Using our non-Hermitian AAH model and its zero modes, a topological lasing mode can be selected via a spatially extended pump, such as the interference pattern formed by two optical pumping beams.  The topological lasing mode can even be enabled or disabled by tuning the phase and period of the pumping pattern.

%% This points to further opportunities of experimental observation of topological protected edge modes in non-Hermitian quasicrystals.

\section{Model}

We consider a one-dimensional chain with coupling $t$ between nearest neighbors and a purely imaginary on-site potential described by a sinuoidal modulation, as depicted in Fig.~\ref{fig:schematic_phase}(a).  The Schr\"odinger equation is
\begin{align}
  \begin{aligned}
    t(\psi_{n+1}+\psi_{n-1}) + iV_n\psi_n = E\psi_n, \\
    V_n = V\sin(2\pi\alpha n+\delta),
  \end{aligned}
  \label{eq:1}
\end{align}
where $\psi_n$ is the wavefunction at site $n$, $E$ is the eigenenergy, and $V$, $\alpha$ and $\delta$ are the amplitude, inverse period, and phase of the potential modulation function.  We will set the unit of energy so that $t = 1$.  We consider rational values of $\alpha=q/p$, where $p$ and $q$ are coprime positive integers; hence, the modulation function is commensurate with the underlying lattice, and the model is periodic with $p$ sites per unit cell \cite{Thouless1982}.

If $p$ is even, the bulk Hamiltonian $\hat{\mathcal{H}}_k$ (a $p \times p$ matrix) satisfies the non-Hermitian particle-hole symmetry \cite{Qi2018}
\begin{equation}
  -\hat{\mathcal{H}}_k = \mathcal{\hat{C}}\hat{T} \hat{\mathcal{H}}^*_{-k} \hat{T}\mathcal{\hat{C}} = \mathcal{\hat{C}}\hat{T} \hat{\mathcal{H}}^\dagger_k \hat{T}\mathcal{\hat{C}},
  \label{eq:CTsym}
\end{equation}
where $\mathcal{\hat{C}} = I_{p/2} \otimes \sigma_{z}$, with $I_{p/2}$ denoting the $p/2\times p/2$ identity matrix and $\sigma_{z}$ denoting the third Pauli matrix, and $\hat{T}$ is the complex conjugation (time-reversal) operator.  Eq.~\eqref{eq:CTsym} implies that the bulk eigenstates either occur in pairs with energies $\{E_1,E_2\}$ satisfying $\hat{E}_1(k) = -\hat{E}^*_2(k)$, or form a flat band with purely imaginary energy \cite{Qi2018}.  Moreover, for a finite lattice with $N$ sites (with $N$ even), the Hamiltonian $\hat{H}$ obeys the non-Hermitian particle-hole symmetry
\begin{equation}
  \{\hat{H}, \hat{C}\hat{T}\} = 0,
\end{equation}
where $\hat{C} = \mathbf{I}_{N/2} \otimes \sigma_{z}$.

We will focus on the case of $p=4\textbf{N}$, where $\textbf{N} \in \mathbb{Z}^+$.  In this case, the bulk bandstructure can host a real line gap, meaning a gap in the real part of the spectrum \cite{Kawabata2019,Bergholtz2021}, around $\mathrm{Re}(E) = 0$.  Such a gap does not appear for other choices of $\alpha$ (see Supplemental Materials \cite{SM}).  As an example, Fig.~\ref{fig:schematic_phase}(b) plots the complex spectrum for $\alpha = 3/8$, using a lattice of $N = 200$ sites with open boundary conditions (OBC). In the bulk spectrum, calculated using PBC with the same lattice parameters, the real line gap closes at $m\pi/4$ where $m\in\mathbb{Z}$.  In Fig.~\ref{fig:schematic_phase}(b), it appears that the gap does not fully close at certain of these points (e.g., at $\delta = \pi/2$), but this is a finite-size effect; for larger $N$, the OBC spectrum has gap-closings at the same points as the PBC spectrum (for details, see the Supplemental Materials \cite{SM}).

Within half of the gaps, the lattice with OBC exhibits eigenenergies with $\mathrm{Re}(E) = 0$, plotted as red curves in Fig.~\ref{fig:schematic_phase}(b).  The wavefunctions of these ``zero modes'' are exponentially localized to the lattice boundary, as shown in Fig.~\ref{fig:schematic_phase}(c)--(d).  The zero modes preserve the non-Hermitian particle-hole symmetry: each eigenvector $|\psi\rangle$ obeys $|\psi\rangle = e^{i\theta}\hat{C}\hat{T}|\psi\rangle$, where $\theta$ is some global phase factor \cite{Ge2017}.  Note also that the zero modes need not have $\mathrm{Im}(E) = 0$; in fact, we see from the lower panel of Fig.~\ref{fig:schematic_phase}(b) that they can have larger $\mathrm{Im}(E)$ than the bulk states.  We will explore the possibility of using this feature for lasing in Section~\ref{sec:lasing}. In the Supplemental Materials \cite{SM}, we show that the ``zero modes'' are robust to the disorders preserving particle-hole symmetry.

In Fig.~\ref{fig:schematic_phase}(b), we can also see some in-gap states in the other bandgaps, away from $\mathrm{Re}(E) = 0$.  These are similar to the topological boundary states of the original AAH model \cite{tknn, Janot1994, Kraus2016, Zilberberg2021}, and are not the focus of the present work.

Takata and Notomi \cite{Takata2018} have studied the case of $p=4$, $q=1$, which corresponds to the repeating gain/loss sequence $\{g_1, -g_2, -g_1, g_2\}$.  In particular, they noted the existence of zero modes induced by the imaginary potential.  The present work extends these results to a wider range of gain/loss modulations based on non-Hermitian AAH models.

\section{Topological phases}

The zero modes introduced in the previous section are linked to topological features of the non-Hermitian bandstructure.  These are expressible using the non-Hermitian Berry connection, calculated via a biorthogonal product instead of the Hermitian inner product \cite{Garrison1988, LiangPRA2013, Lieu2018}.

The non-Hermitian band topology can be characterized in two complementary ways \cite{SM}. The first approach involves the non-Hermitian generalization \cite{Luo2019,LeePRB2020,Taberner2022,Hu2022} of the electric polarization \cite{Smith1993,Resta1998}.  When there is a real line gap, we can calculate the non-Abelian, non-Hermitian Berry connection for all bands with $\mathrm{Re}(E) < 0$, and use the nested Wilson loop method \cite{BenalcazarPRB2017,BenalcazarScience2017} to integrate it around the Brillouin zone.  This procedure has previously been shown to yield quantized polarizations in other non-Hermitian systems with real line gaps, e.g.~non-Hermitian higher-order topological insulators \cite{Luo2019}.  The second approach to characterizing the band topology is the global Berry phase \cite{LiangPRA2013, Parto2018, Takata2018}, which involves integrating the non-Hermitian Berry connections for all bands (with care taken to fix the gauge and sort the bands \cite{Wagner2017, Comaron2020, SM}).  Both methods are based on the bulk band structure, derived under PBC.

\begin{figure}
  \centering
  \includegraphics[width=0.48\textwidth]{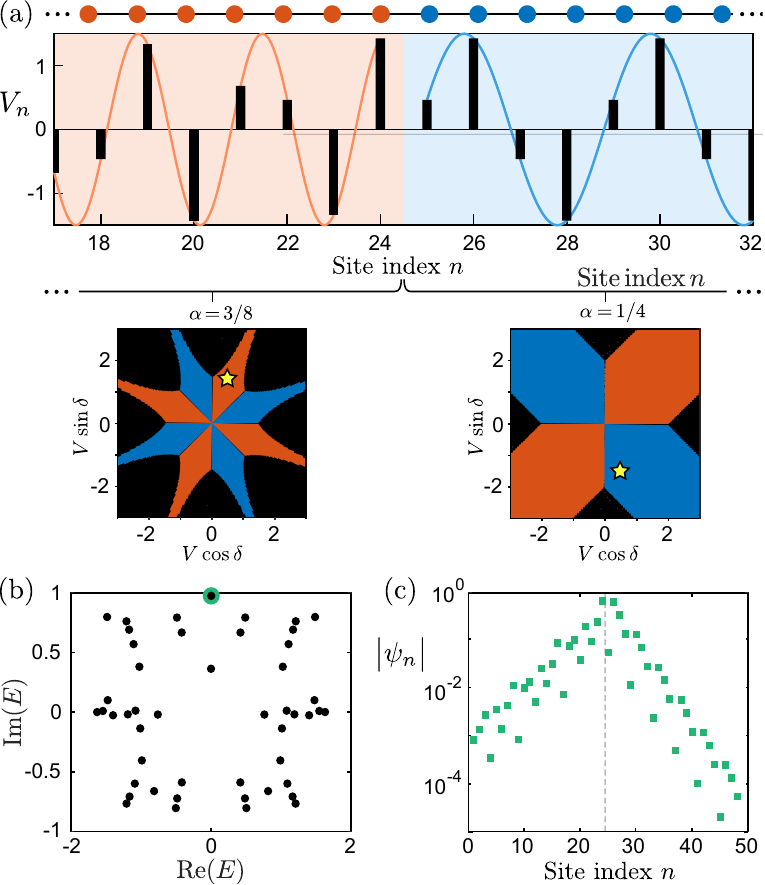}
  \caption{(a) Lattice formed by joining two chains with different gain/loss modulations.  In the upper panel, the left (right) domain, marked in orange (blue), is topologically nontrivial (trivial).  In the middle panel, the black bars indicate $V_n$, the gain/loss on site $n$, and the solid curves plot the modulation functions, which notably have different periods in the two domains.  For the left domain, $\alpha = 3/8$, $V=1.5$, and $\delta = 0.4\pi$; for the right domain, $\alpha=1/4$, $V=1.5$, and $\delta=-0.4\pi$.  In the lower panels, the phase diagrams for the two domains are shown, with the choice of modulation parameters marked by yellow stars.  (b) Complex eigenenergy spectrum for the lattice, with a total of $N=48$ sites (24 in each domain).  The mirror symmetry around $\mathrm{Re}(E) = 0$ is due to the non-Hermitian particle-hole symmetry in Eq.~\eqref{eq:CTsym}.  (c) Spatial distribution of the zero mode highlighted in green in (b).  Vertical dashes indicate the domain wall.  }
  \label{fig:domain_wall}
\end{figure}

When there is a real line gap at $\mathrm{Re}(E) = 0$, the polarization and global Berry phase calculations are in agreement, and yield the topological phase diagrams shown in Fig.~\ref{fig:schematic_phase}(e).  These phase diagrams are plotted using the modulation parameters $(V,\delta)$ as polar coordinates, for various $\alpha = q/p$ with $p$, $q$ coprime and $p$ a multiple of 4. In the orange regions, the bandstructure gives quantized polarization $p_x=1/2$ and a global Berry phase of $2\pi$.  In the blue regions, the polarization and global Berry phase vanish.  In the black regions, there is no real line gap at $\mathrm{Re}(E) = 0$ and the polarization calculation is inapplicable; we will discuss the lattice's behavior in this regime later in this section.  Evidently, the real line gap phases form $p$ spokes in the phase diagram, extending outward from the origin $V = 0$, and alternating between trivial and nontrivial phases.  The phase diagrams for other $\alpha = q/p$ are consistent with the pattern shown in Fig.~\ref{fig:schematic_phase}(e).

\begin{figure}
  \centering
  \includegraphics[width=0.48\textwidth]{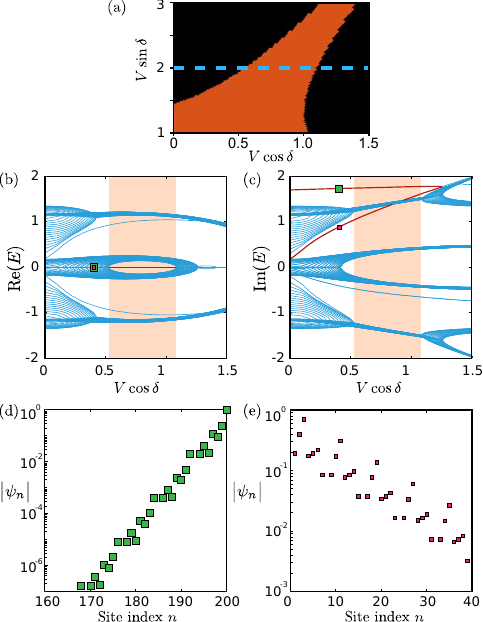}
  \caption{(a) Close-up view of the phase diagram for $\alpha=3/8$.  Dashes indicate a trajectory corresponding to $V \sin\delta=2$.  (b)--(c) Complex eigenenergy spectrum corresponding to the dashes in (a), for lattice size $N=200$.  The ranges corresponding to the nontrivial real line gap phase of bulk spectrum are highlighted in orange.  The topological zero modes, plotted in red, persist even when the real line gap in bulk spectrum closes, and remain exponentially localized to the boundary.  (d)--(e)  Spatial distribution of the zero modes at $V\cos\delta = 0.4$, marked by squares in (b)--(c).}
  \label{fig:phase_transition}
\end{figure}

To test whether the phase diagrams correctly predict the existence of zero modes, we examine the behavior at domain walls between different modulation functions \cite{Kraus2012, Verbin2013, Verbin2015, Takata2018, Ota2018, Lang2018}.  The lattice shown in Fig.~\ref{fig:domain_wall}(a) consists of two adjacent domains with different gain/loss distributions.  The two modulation functions have different $\alpha$ ($3/8$ and $1/4$), as well as different $\delta$, as indicated by the phase diagrams in the lower panels of Fig.~\ref{fig:domain_wall}(a).  With the two domains chosen to be topologically inequivalent, we see that the complex spectrum, plotted Fig.~\ref{fig:domain_wall}(b), contains a zero mode (highlighted in green).  Its wavefunction is exponentially localized to the domain wall, as shown in Fig.~\ref{fig:domain_wall}(c).  (Note that this zero mode has the largest $\mathrm{Im}(E)$ among all the eigenstates; we will discuss the significance of this in Sec.~\ref{sec:lasing}.  The other zero mode that can be seen in Fig.~\ref{fig:domain_wall}(b) is localized to the opposite end of the topologically nontrivial domain, rather than the domain wall.)  In the Supplemental Materials \cite{SM}, we show other combinations of modulation parameters, which all behave as expected.  In particular, if the domains are both trivial or both nontrivial, there is no zero mode at the domain wall.  This verifies that the zero modes arise from a non-Hermitian topological bulk-edge correspondence.

An interesting feature of the non-Hermitian zero modes is that they can persist for a short but nonzero interval after the closing of the real line gap in bulk spectrum, pinnted to $\mathrm{Re}(E) = 0$. This contrasts with the Hermitian case, where the closing of the band gap causes zero modes and other localized boundary states to hybridize with bulk states and lose their localized character.  In Fig.~\ref{fig:phase_transition}(a), we plot a parametric trajectories in the $\alpha = 3/8$ phase diagram, extending into the gapless (i.e., no real line gap in bulk spectrum) phases to each side of the gapped phase.  The complex band energies along these trajectories are plotted in Fig.~\ref{fig:phase_transition}(b)--(c).  When the real line gap in bulk spectrum closes, the complex-valued zero mode energies and bulk energy bands (specifically, their imaginary parts) do not overlap.   Hence, the zero modes remain spatially localized, as shown in Fig.~\ref{fig:phase_transition}(d)--(e). In the Supplemental Materials \cite{SM}, we show that zero modes vanish by coalescing with each other at exceptional points \cite{Ozdemir2019, Ashida2020}, rather than hybridizing with bulk states; moreover, within the gapless phase, they are robust against disorder that preserves particle-hole symmetry. Related behavior has recently been pointed out in the context of topological crystalline insulators, where higher-order topological modes can persist despite having $\mathrm{Re}(E)$ degenerate with the bulk bands \cite{BenalcazarPRB2020, CerjanPRL2020}.

\section{Mode Selection in Laser Arrays}
\label{sec:lasing}

\begin{figure}
  \centering
  \includegraphics[width=0.48\textwidth]{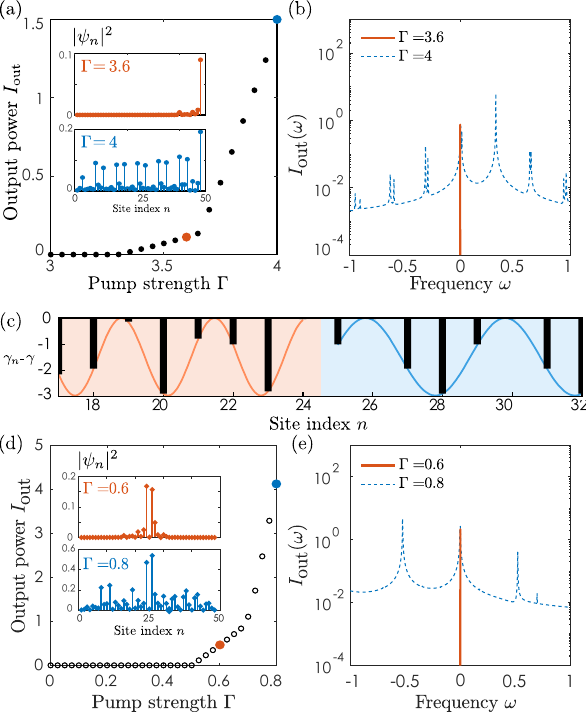}
  \caption{(a)--(b) Simulation results for a laser with a spatially modulated pump strength, given by Eq.~\eqref{gainsat_1}.  The lattice consists of a single domain with $\alpha=3/8$, $\delta=0.4\pi$, and size $N = 48$, with passive loss rate $\gamma=3$.  (a) ﻿Output intensity $I_{\mathrm{out}}$ versus pump strength $\Gamma$. Inset: spatial distribution of $|\psi_n|^2$ at $t=5000$ for $\Gamma=3.6$ (top) and $\Gamma=4$ (bottom). (b) Output spectrum for $\Gamma=3.6$ (orange line) and $\Gamma=4$ (blue dashes). (c)--(e) Simulation results for a laser with spatially modulated loss and uniform pump strength, given by Eq.~\eqref{gainsat_2}, with lattice size $N = 48$.  (c) Spatial distribution of the loss term $\gamma_n-\gamma$ (black bars), along with the underlying modulation functions.  These modulation functions form a domain wall in the center of the lattice, and are identical to those of Fig.~\ref{fig:domain_wall} up to a constant offset.  (d) ﻿Output intensity $I_{\mathrm{out}}$ versus pump strength $\Gamma$.  Inset: spatial distribution of $|\psi_n|^2$ at $t=5000$ for pump strengths $\Gamma=0.6$ (top) and $\Gamma=0.8$ (bottom). (d) Output spectrum at $\Gamma = 0.6$ (orange line) and $\Gamma = 0.8$ (blue dashes).}
  \label{fig:lasing}
\end{figure}

The non-Hermitian AAH model can be used as the basis for a topological laser distinct from the other topological lasers studied to date \cite{StJean2017, Ota2018, Parto2018, Zhao2018, Harari2018, Bandres2018, ZengYQ2020, ZYang2020, Harder2021, Dikopoltsev2021, Choi2021, Comaron2020}.  It has previously been noted that lasers are a natural setting for realizing and exploiting non-Hermitian topological phenomena, since they necessarily contain gain (stimulated emission) and loss (outcoupling and material dissipation).  Thus, for instance, researchers have implemented laser arrays based on the PT symmetric SSH model, with lasing modes based on the non-Hermitian zero modes of that model \cite{Parto2018, Zhao2018}.

In this context, the non-Hermitian AAH model's most striking feature is that its properties are governed directly by the gain/loss modulation function, whose period differs from (and can be signficantly larger than) that of the photonic lattice.  One interesting possibility is to excite the lattice using a sinuoidally varying pump profile, such as an interference pattern of two optical pumping beams.  The period and phase of the pump profile could be easily varied to access different parts of the non-Hermitian AAH model's phase diagram.

To investigate this, we consider a laser model consisting of a non-Hermitian AAH chain with a \textit{nonlinear} imaginary potential $iV_n$, where
\begin{align}
  \begin{aligned}
    V_{n} &= \frac{\Gamma\lambda_n}{1+|\psi_{n}|^2} - \gamma, \\
    \lambda_n &= \frac{1}{2}\big[1 + \sin(2\pi\alpha n+\delta)\big].
  \end{aligned}
  \label{gainsat_1}
\end{align}
Here, $\Gamma$ is an overall pump strength, $\lambda_n \in [0,1]$ is the spatial modulation of the pump strength, $|\psi_n|^2$ is the local intensity on site $n$, and $\gamma$ is a passive loss rate (which can include outcoupling loss).  The $1+|\psi_n|^2$ denominator represents the effects of gain saturation \cite{Ge2017, Harari2018}.  Taking $\alpha=3/8$, $\delta=0.4\pi$, and $\gamma=3$, we performed time-domain simulations by numerically integrating the nonlinear equation $i\partial_t|\psi\rangle = H(|\psi\rangle) |\psi\rangle$ \cite{Ge2017, Harari2018, Yang2020}.  The wavefunction is initialized to the random values $\psi_{n}(t=0) = (\alpha_{n} + i \beta_{n})f_0$, where $\alpha_{n}, \beta_{n}$ are drawn independently from the standard normal distribution, $n$ is the site index and $f_0 = 0.01$ is a scale factor.  The simulated time interval is $t \in [0, 5000]$, long enough for transient oscillations to cease.  The laser output $I_{\mathrm{out}}$ is obtained by averaging the on-site intensities $|\psi_{n}|^2$ over the evolution interval of $t \in [2000, 5000]$ (we assume equal outcoupling from each site, with normalized power units).

The resulting plot of $I_{\mathrm{out}}$ versus pump strength $\Gamma$ is shown in Fig.~\ref{fig:lasing}(a).  Because the zero mode of the linear lattice has (in this case) the highest relative gain, it lases first.  As shown in the inset of Fig.~\ref{fig:lasing}(a), the frequency spectrum at $\Gamma = 3.6$ consists of a single peak at $\omega = 0$.  (Note that in an actual laser, $\omega$ is a frequency detuning, relative to the natural frequency of the decoupled resonators.)  The intensity is localized to one boundary of the lattice, as shown in the upper panel of inset in Fig.~\ref{fig:lasing}(a).  These two features of the lasing mode---the pinning of the frequency to $\omega = 0$ and the spatial localization---are inherited from the linear non-Hermitian AAH model, and are selected by the choice of the pump's spatial modulation $\lambda_n$, which can be easily adjusted (e.g., by changing the interference pattern of an optical pump). When $\Gamma$ is further increased, the additional modes of the lattice also start to lase \cite{SM}, and the system enters the multi-mode lasing regime, as shown for the case of $\Gamma = 4$ in Fig.~\ref{fig:lasing}(b) and the bottom panel of inset in Fig.~\ref{fig:lasing}(a).

An alternative way to access the non-Hermitian AAH model with a laser array is to use loss engineering.  We can modulate the (linear) loss on individual sites, and then pump the entire lattice, as described by the imaginary potential $iV_n$, where
\begin{align}
  V_{n} = \frac{\Gamma}{1+|\psi_{n}|^2} + \gamma_n - \gamma,
  \label{gainsat_2}
\end{align}
For this case, we suppose $\gamma_n - \gamma$ is formed by two modulation functions with a domain wall at the center, as shown in Fig.~\ref{fig:lasing}(c).  Within each domain, $\gamma_n = V\sin(2\pi\alpha n+\delta)$, and we pick the same values of $V$, $\alpha$ and $\delta$ as in Fig.~\ref{fig:domain_wall}.  We also include an additional constant loss $\gamma = \mathrm{Im}(E_0) + 0.5$, where $E_0$ is the eigenenergy of the desired zero mode [square marker in Fig.~\ref{fig:domain_wall}(b)].  As shown in Fig.~\ref{fig:lasing}(d), this sets the laser threshold to $\Gamma = 0.5$.  For some range of pump strengths above threshold, the lasing frequency is pinned to $\omega = 0$ and localized at the domain wall, as shown in Fig.~\ref{fig:lasing}(e) and the upper panel of Fig.~\ref{fig:lasing}(d) for pump strength $\Gamma = 0.6$. Multi-mode lasing is observed at higher pump strengths (e.g., $\Gamma = 0.8$). Although the spatial modulation in lattice recalls the conventional  distributed-feedback (DFB) laser, the basic mechanism is quite different. As the spatial modulation in DFB lasers only provide reflection (optical feedback), while the imaginary  AAH-style potential alters the topology properties and give rise to topologically protected edge modes. The laser design proposed here also differs from previous 1D topological lasers \cite{Zhao2018, Parto2018}, which were based on the SSH model and its non-Hermitian variants.  In particular, the topological modes of the SSH lasers are inherited from the Hermitian SSH model, whereas the present topological phases are generated from non-Hermitian gain/loss modulations.

\section{Discussion}

We have demonstrated that a non-Hermitian variant of the AAH model, consisting of a sinusoidally-modulated potential that is not real but rather purely imaginary, can exhibit topological boundary modes with purely imaginary energy.  These zero modes are found when the modulation period is a multiple of four lattice constants, and are pinned to $\mathrm{Re}(E) = 0$ by an unbroken non-Hermitian particle-hole symmetry \cite{Ge2017}.  Our results generalize the period-four lattice found by Takata and Notomi \cite{Takata2018}
to a wider family of AAH-type imaginary potentials.  The complex bandstructure has two distinct phases with real line gaps at $\mathrm{Re}(E) = 0$, characterized by non-Hermitian topological invariants.  The invariants correctly predict the existence of zero modes, even for domain walls between modulations with different periods.

Previously, Hermitian zero modes have been observed in a variant of AAH model that has commensurate modulations in the hoppings (rather than the on-site potential) \cite{Ganeshan2013}.  However, that model was based on a Hermitian particle-hole symmetry different from Eq.~\eqref{eq:CTsym}, and we have not found any deeper relationship between these sets of results.

In the non-Hermitian AAH model, it is possible to tune the imaginary potential  so that the zero modes can have the highest relative gain of all the eigenstates.  This property can be exploited for mode selection in laser arrays, as we showed using simulations.  One interesting possibility is to use optical pumping beams in an interference pattern (corresponding to spatially modulated gain) to control the lasing of the zero modes; alternatively, one can modulate the loss in the laser array and pump uniformly.  In both cases, our simulations results show that a non-Hermitian zero mode can be the first lasing mode, and retain its key characteristics (frequency pinning and spatial localization) from the lasing threshold up to the onset of multi-mode lasing.

This work was supported by the Singapore MOE Academic Research Fund Tier 3 Grant MOE2016-T3-1-006, Tier 2 Grant MOE2019-T2-2-085, Tier 1 Grant RG148/20, and Singapore National Research Foundation (NRF) Competitive Research Program (CRP) NRF-CRP18-2017-02 and NRF-CRP23-2019-0007; Li-Jun Lang was supported by the National Natural Science Foundation of China (Grant No.~11904109), the Guangdong Basic and Applied Basic Research Foundation (Grant No.~2019A1515111101).


\begin{thebibliography}{99}

\bibitem{Harper1955}
  P.~G.~Harper, 
  Single Band Motion of Conduction Electrons in a Uniform Magnetic Field,
  Proc.~Phys.~Soc.~Sec.~A \textbf{68}, 874 (1955).

\bibitem{Aubry1980} 
  S.~Aubry and G.~Andr\'e,
  Analyticity breaking and Anderson localization in incommensurate lattices,
  Ann~ Isr.~Phys.~Soc.~\textbf{3}, 133 (1980).  


\bibitem{Thouless1982}
  D.~J.~Thouless, M.~Kohmoto, M.~P.~Nightingale, and M.~den Nijs,
  Quantized Hall Conductance in a Two-Dimensional Periodic Potential,
  Phys.~Rev.~Lett.~\textbf{49}, 405 (1982). 

%% quasicrystal review

\bibitem{tknn} D.~Thouless, M.~Kohmoto, M.~P.~Nightingale, and M.~den
  Nijs, Quantized Hall Conductance in a Two-Dimensional Periodic Potential, Phys.~Rev.~Lett.~{\bf 49}, 405 (1982).

\bibitem{Janot1994}
  J.~Faist,
  Quasicrystals,~1st ed. (Oxford University Press, Oxford, 1994).    

\bibitem{Kraus2016}
  Y.~E.~Kraus and O.~Zilberberg,
  Quasiperiodicity and topology transcend dimensions,
  Nat.~Phys.~\textbf{9}, 981 (2018).  
  
\bibitem{Zilberberg2021}
  O.~Zilberberg,
  Topology in quasicrystals,
  Opt.~Mat.~Express~\textbf{11}, 1143 (2021). 
  
%% AAH with PH symmetry

\bibitem{Ganeshan2013}
  S.~Ganeshan, K.~Sun, and S.~D.~Sarma, 
  Topological Zero-Energy Modes in Gapless Commensurate Aubry-Andr\'e-Harper Models,
  Phys.~Rev.~Lett.~\textbf{110}, 180403 (2013).

\bibitem{Ryu2002}
  S.~Ryu and Y.~Hatsugai,
  Topological Origin of Zero-Energy Edge States in Particle-Hole Symmetric Systems,
  Phys.~Rev.~Lett.~\textbf{89}, 077002 (2002). 

\bibitem{Kitaev2001}
  A. Y. Kitaev, Unpaired Majorana fermions in quantum wires, Phys.-Usp. 44 (2001).

%% non-Hermitian AAH models 

\bibitem{Longhi2014}   % uniformly distributed PT-symmetric gain/loss
  S.~Longhi,
  PT-symmetric optical superlattices,
  J.~Phys.~A~\textbf{47}, 165302 (2014).  

\bibitem{Yuce2014}   % uniformly distributed PT-symmetric gain/loss
  C.~Yuce,
  PT-symmetric Aubry-Andr\'e model,
  Phys. Lett.~A~\textbf{378}, 2024 (2014).  

\bibitem{Harter2016}   % locally distributed PT-symmetric gain/loss
  A.~K.~Harter, T.~E.~Lee, and Y.~N.~Joglekar,
  PT-breaking threshold in spatially asymmetric Aubry-Andr`e and Harper models: Hidden symmetry and topological states,
  Phys.~Rev.~A~\textbf{93}, 062101 (2016). 

\bibitem{Longhi2019}   % uniformly distributed PT-symmetric gain/loss
  S.~Longhi,
  Topological Phase Transition in non-Hermitian Quasicrystals,
  Phys.~Rev.~Lett.~\textbf{122}, 237601 (2019). 

\bibitem{Jiang2019}    % non-reciprocal hopping
  H.~Jiang, L.~Lang, C.~Yang, S.~Zhu, and S.~Chen,
  Interplay of non-Hermitian skin effects and Anderson localization in nonreciprocal quasiperiodic lattices,
  Phys.~Rev.~B~\textbf{100}, 054301 (2019). 

\bibitem{Zeng2020}    % mobility edge
  Q.~Zeng, Y.~Yang, and R.~L\"u,
  Topological phases in one-dimensional nonreciprocal superlattices,
  Phys.~Rev.~B~\textbf{101}, 125418 (2020). 

\bibitem{TLiu2020}    % mobility edge
  T.~Liu, H.~Guo, Y.~Pu, and S.~Longhi,
  Generalized Aubry-Andr\'e self-duality and mobility edges in non-Hermitian quasiperiodic lattices,
  Phys.~Rev.~B~\textbf{102}, 024205 (2020). 

\bibitem{YLiu2020}    % mobility edge
  Y.~Liu, X.~P.~Jiang, J.~Cao, and S.~Chen,
  Non-Hermitian mobility edges in one-dimensional quasicrystals with parity-time symmetry,
  Phys.~Rev.~B~\textbf{101}, 174205 (2020). 

\bibitem{Longhi2021}    % non-reciprocal hopping
  S.~Longhi,
  Phase transitions in a non-Hermitian Aubry-Andr\'e-Harper model,
  Phys.~Rev.~B~\textbf{103}, 054203 (2021). 

\bibitem{Weidemann2022}    % Floquet modulations, non-reciprocal hopping
 S.~Weidemann, M.~Kremer, S.~Longhi, and A.~Szameit,
 Topological triple phase transition in non-Hermitian Floquet quasicrystals,
 Nature~\textbf{601}, 354 (2022).
 
%% Non-Hermitian reviews

\bibitem{Ozdemir2019}
  {\c{S}}.~K.~{\"O}zdemir, S.~Rotter, F.~Nori, and L.~Yang,
  Parity-time symmetry and exceptional points in photonics,
  Nat.~Mater.~\textbf{18}, 783 (2019).  

\bibitem{Kawabata2019}
  K.~Kawabata, K.~Shiozaki, M.~Ueda, and M.~Sato, 
  Symmetry and Topology in Non-Hermitian Physics,
  Phys.~Rev.~X~\textbf{9}, 041015 (2019).  

\bibitem{Ashida2020}
  Y.~Ashida, Z.~Gong, and M.~Ueda,
  Non-Hermitian Physics,
  Adv.~Phys.~\textbf{69}, 249 (2020).

%% PT SSH
  
\bibitem{Schomerus2013}
  H.~Schomerus,
  Topologically protected midgap states in complex photonic lattices,
  Opt.~Lett.~\textbf{38}, 1912 (2013).

\bibitem{Lang2018}
  L.~Lang, Y.~Wang, H.~Wang and Y.~D.~Chong,
  Effects of non-Hermiticity on Su-Schrieffer-Heeger defect states,
  Phys.~Rev.~B~\textbf{98}, 094307 (2018).   

\bibitem{Pan2018}
  M.~Pan, H.~Zhao, P.~Miao, S.~Longhi and L.~Feng, 
  Photonic zero mode in a non-Hermitian photonic lattice,
  Nat.~Commun.~\textbf{9}, 1308 (2018).  

%% PT lasers based on spatial gain/loss distribution

\bibitem{Parto2018}
  M.~Parto, S.~Wittek, H.~Hodaei, G.~Harari, M.~A.~Bandres, J.~Ren, M.~C.~Rechtsman, M.~Segev, D.~N. ~Christodoulides and M.~Khajavikhan,
  Edge-Mode Lasing in 1D Topological Active Arrays,
  Phys.~Rev.~Lett.~\textbf{120}, 113901 (2018).
  
\bibitem{Zhao2018}
  H.~Zhao, P.~Miao, M.~H.~Teimourpour, S.~Malzard, R.~E.~Ganainy, H.~Schomerus and L.~Feng,
  Topological hybrid silicon microlasers,
  Nat.~Commun.~\textbf{9}, 981 (2018).  

%% NHPH symmetry

\bibitem{Ge2017}  
  L.~Ge,
  Symmetry-protected zero-mode laser with a tunable spatial profile,
  Phys.~Rev.~A~\textbf{95}, 023812 (2017).
 
%% p=2 (N=1) case %%
\bibitem{Qi2018}
  B.~Qi, L.~Zhang, and L.~Ge,
  Defect States Emerging from a Non-Hermitian Flatband of Photonic Zero Modes,
  Phys.~Rev.~Lett.~\textbf{120}, 093901 (2018). 

%% Intrinsically non-Hermitian boundary states
\bibitem{Lee2016}
  T.~E.~Lee,
  Anomalous Edge State in a Non-Hermitian Lattice,
  Phys.~Rev.~Lett.~\textbf{116}, 133903 (2016).  
  
\bibitem{Leykam2017}
  D.~Leykam, K.~Y.~Bliokh, C.~Huang, Y.~D.~Chong, and F.~Nori,
  Edge Modes, Degeneracies, and Topological Numbers in Non-Hermitian Systems,
  Phys.~Rev.~Lett.~\textbf{118}, 040401 (2017).

%% p=4 (N=2) case %%

\bibitem{Takata2018}
  K.~Takata and M.~Notomi,
  Photonic Topological Insulating Phase Induced Solely by Gain and Loss,
  Phys.~Rev.~Lett.~\textbf{121}, 213902 (2018). 

%% electric polarization in non-Hermitian system %%

\bibitem{Luo2019} 
  X.~Luo and C.~Zhang,
  Higher-Order Topological Corner States Induced by Gain and Loss, 
  Phys.~Rev.~Lett.~\textbf{123}, 073601 (2019).

\bibitem{LeePRB2020} 
  E.~Lee, H.~Lee, and B.-J.~Yang,
  Many-body approach to non-hermitian physics in fermionic systems, 
  Phys.~Rev.~B~\textbf{101}, 121109 (2020).

\bibitem{Taberner2022} 
  C.~Ortega-Taberner, L.~R{\o}dland, and M.~Hermanns,
  Polarization and entanglement spectrum in non-hermitian systems, 
  Phys.~Rev.~B~\textbf{105}, 075103 (2022).
  
%\bibitem{Hu2022} %% electric polarization in non-Hermitian system with CT symmetry %%
%  J.~Hu, C.~A.~Perroni, G.~D.~Filippis, S.~Zhuang, L.~Marrucci, and F.~Cardano,
%  Electric polarization and its quantization in one-dimensional non-Hermitian chains,
%  arXiv preprint arXiv:2203.11902 (2022).
  
\bibitem{Hu2022} 
  J.~Hu, C.~A.~Perroni, G.~D.~Filippis, S.~Zhuang, L.~Marrucci, and F.~Cardano,
  Electric polarization and its quantization in one-dimensional non-Hermitian chains, 
  Phys.~Rev.~B~\textbf{107}, L121101 (2023).  

%% 1D topological lasers based on partial pump scheme

\bibitem{StJean2017}
  P.~St-Jean, V.~Goblot, E.~Galopin, A.~Lema{\^\i}tre, T.~Ozawa, L.~Le Gratiet, I.~Sagnes, J.~Bloch, and A.~Amo,
  Lasing in topological edge states of a one-dimensional lattice,
  Nat.~Photon.~\textbf{11}, 651 (2017).

\bibitem{Ota2018}          %  PC, cavity mode lasing
  Y.~Ota, R.~Katsumi, K.~Watanabe, S.~Iwamoto and Y.~Arakawa,
  Topological photonic crystal nanocavity laser,
  Commun.~Phys.~\textbf{1}, 1 (2018)
  
%% 2D topological lasers based on partial pump scheme
 
\bibitem{Harari2018}
  G.~Harari, Miguel.~A.~Bandres, Y.~Lumer, M.~C.~Rechtsman, Y.~D.~Chong, M.~Khajavikhan,
  D.~N.~Christodoulides, and M.~Segev,
  Topological insulator laser: Theory,
  Science~\textbf{359}, eaar4003 (2018). 

\bibitem{Bandres2018}
  M.~A.~Bandres, S.~Wittek, G.~Harari, M.~Parto, J.~Ren, M.~Segev, D.~N.~Christodoulides, and M.~Khajavikhan,
  Topological insulator laser: Experiments,
  Science~\textbf{359}, eaar4005 (2018).

\bibitem{ZengYQ2020}
  Y.~Zeng, U.~Chattopadhyay, B.~Zhu, B.~Qiang, J.~ Li, Y.~Jin, L.~Li, A.~G.~Davies, E.~H.~Linfield, B.~Zhang, Y.~D.~Chong, and Q.~J.~Wang,
  Electrically pumped topological laser with valley edge modes,
  Nature~\textbf{578}, 246 (2020).  

\bibitem{ZYang2020}
  Z.~Yang, Z.~Shao, H.~Chen, X.~Mao, and R.~Ma,
  Spin-momentum-locked edge mode for topological vortex lasing,
  Phys.~Rev.~Lett.~\textbf{125}, 013903 (2020). 

\bibitem{Harder2021}
  T.~H.~Harder, M.~Sun, O.~A.~Egorov, I.~Vakulchyk, J.~Beierlein, P.~Gagel, M.~Emmerling, C.~Schneider, U.~Peschel, S.~Klembt, and S.~Hofling,
  Coherent topological polariton laser,
  ACS~Photonics~\textbf{8}, 1377 (2021).  
  
\bibitem{Dikopoltsev2021}
  A.~Dikopoltsev, T.~H.~Harder, E.~Lustig, O.~A.~Egorov, J.~Beierlein, A.~Wolf, Y.~Lumer, M.~Emmerling, C.~Schneider, S.~Höfling, M.~Segev, and S.~Klembt, 
  Topological insulator vertical-cavity laser array, 
  Science~\textbf{373}, 1514 (2021).

\bibitem{Choi2021}
  J.~H.~Choi, W.~E.~Hayenga, Y.~G.~Liu, M.~Parto, B.~Bahari, D.~N.~Christodoulides, and M.~Khajavikhan, 
  Room temperature electrically pumped topological insulator lasers,
  Nat.~Commun.~\textbf{12}, 1 (2021).  
  
\bibitem{Qi2019}
  B.~Qi, H.~Chen, L.~Ge, P.~Berini, and R.~Ma,
  Parity-Time Symmetry Synthetic Lasers: Physics and Devices,  
  Adv.~Opt.~Mater.~\textbf{7}, 1900694 (2019). 

%% non-Hermitian review   

\bibitem{Bergholtz2021}
  E.~J.~Bergholtz, J.~C.~Budich, F.~K.~Kunst,
  Exceptional topology of non-Hermitian systems,  
  Rev.~Mod.~Phys.~\textbf{93}, 015005 (2021). 

%% non-Hermitian Berry phase       

\bibitem{Garrison1988}
   J.~C.~Garrison and E.~M.~Wright,
  Complex geometrical phases for dissipative systems,
  Phys.~Lett.~A \textbf{128}, 177 (1988).

\bibitem{LiangPRA2013}
  S.~D.~Liang and G.~Y.~Huang,
  Topological invariance and global Berry phase in non-Hermitian systems,
  Phys.~Rev.~A~\textbf{87}, 012118 (2013). 

%% bi-orthogonal normalization %%

\bibitem{Lieu2018} 
  S.~Lieu,
  Topological phases in the non-Hermitian Su-Schrieffer-Heeger model, 
  Phys.~Rev.~B~\textbf{97}, 045106 (2018).

%% electric polarization in Hermitian system %%

\bibitem{Smith1993} 
  R.~King-Smith and D.~Vanderbilt,
  Theory of polarization of crystalline solids, 
  Phys.~Rev.~B~\textbf{47}, 1651 (1993).
  
\bibitem{Resta1998} 
  R.~Resta,
  Quantum-mechanical position operator in extended systems, 
  Phys.~Rev.~Lett.~\textbf{80}, 1800 (1998).

\bibitem{BenalcazarPRB2017} 
  W.~A.~Benalcazar, B.~A.~Bernevig, and T.~L.~Hughes,
  Electric multipole moments, topological multipole moment pumping, and chiral hinge states in crystalline insulators, 
  Phys.~Rev.~B~\textbf{96}, 245115 (2017).

\bibitem{BenalcazarScience2017} 
  W.~A.~Benalcazar, B.~A.~Bernevig, and T.~L.~Hughes,
  Quantized electric multipole insulators, 
  Science~\textbf{357}, 61 (2017).

\bibitem{Blanco2020}
  M.~Blanco~de~Paz, C.~Devescovi, G.~Giedke, J.~J.~Saenz, M.~G.~Vergniory, B.~Bradlyn, D.~Bercioux and A.~Garci{\'\i}a~Etxarri,
  Tutorial: Computing topological invariants in 2D photonic crystals,
  Adv.~Quantum~Technol.~\textbf{3}, 1900117 (2020).
 
%% global Berry phase in non-Hermitian system %%

\bibitem{Comaron2020}
  P.~Comaron , V.~Shahnazaryan, W.~Brzezicki, T.~Hyart, and M.~Matuszewski,
  Non-Hermitian topological end-mode lasing in polariton systems,
  Phys.~Rev.~Res.~\textbf{2}, 022051(R) (2020). 

\bibitem{Wagner2017}  
  M.~Wagner, F.~Dangel, H.~Cartarius, J.~Main, and G.~Wunner,
  Numerical calculation of the complex berry phase in non-Hermitian systems,
  Acta.~Polytechnica~\textbf{57}, 470 (2017). 

%% SM

\bibitem{SM}
  \textcolor{blue}{See online Supplemental Materials [...]}
   
%% domain wall interface state %% 
  
\bibitem{Kraus2012} 
  Y.~E.~Kraus, Y.~Lahini, Z.~Ringel, M.~Verbin, and O.~Zilberberg,
  Topological States and Adiabatic Pumping in Quasicrystals,
  Phys.~Rev.~Lett.~\textbf{109}, 106402 (2012).
    
\bibitem{Verbin2013} 
  M.~Verbin, O.~Zilberberg, Y.~E.~Kraus, Y.~Lahini, and Y.~Silberberg,
  Observation of Topological Phase Transitions in Photonic Quasicrystals,
  Phys.~Rev.~Lett.~\textbf{110}, 076403 (2013).
  
\bibitem{Verbin2015}
  M.~Verbin, O.~Zilberberg, Y.~Lahini, Y.~E.~Kraus, and Y.~Silberberg,
  Topological pumping over a photonic Fibonacci quasicrystal,
  Phys.~Rev.~B~\textbf{91}, 064201 (2015).  

%% BICs %% 

\bibitem{BenalcazarPRB2020} 
  W.~A.~Benalcazar and A.~Cerjan,
  Bound states in the continuum of higher-order topological insulators, 
  Phys.~Rev.~B~\textbf{101}, 161116(R) (2020).
  
\bibitem{CerjanPRL2020} 
  A.~Cerjan, M.~J\"{u}rgensen, W.~A.~Benalcazar, S.~Mukherjee, and Mikael C. Rechtsman,
  Observation of a Higher-Order Topological Bound State in the Continuum, 
  Phys.~Rev.~Lett.~\textbf{125}, 213901 (2020).
  
%% laser dynamics analysis  

\bibitem{Yang2020}   % mode lock + pulse laser
  Z.~Yang, E.~Lustig, G.~Harari, Y.~Plotnik, Y.~Lumer, M.~A.~Bandres and M.~Segev,
  Mode-Locked Topological Insulator Laser Utilizing Synthetic Dimensions,
  Phys.~Rev.~X~\textbf{10}, 011059 (2020)

  
\end{thebibliography}
\end{document}